\documentclass[reprint,amsmath,amssymb,aps,prd,showpacs,twocolumn]{revtex4-1}

\usepackage{epsfig}
\usepackage{dcolumn}
\usepackage{bm}
\usepackage{tikz}
\usepackage{graphicx, graphics, color}
\usepackage{dcolumn}
\usepackage{bm}
\usepackage{hyperref}
\usepackage[mathlines]{lineno}

\newcommand{\be}{\begin{equation}}
\newcommand{\ee}{\end{equation}}
\newcommand{\ben}{\begin{eqnarray}}
\newcommand{\een}{\end{eqnarray}}

\newcommand{\cM}{{\cal M}}

\newcommand{\p}{\partial}
\newcommand{\na}{\nabla}

\newcommand{\tpsi}{\tilde \psi}

\newcommand{\talpha}{{\tilde \alpha}}

\newcommand{\BK}[1]{\textcolor{blue}{#1}}

\begin{document}


\title{Influence of dark matter on black hole scalar hair}
\author{Bartlomiej Kiczek} 
\email{bkiczek@kft.umcs.lublin.pl}
\author{Marek Rogatko} 
\email{rogat@kft.umcs.lublin.pl}
\affiliation{Institute of Physics, 
Maria Curie-Sklodowska University, 
20-031 Lublin, pl.~Marii Curie-Sklodowskiej 1, Poland}

\date{\today}

\begin{abstract}
Searches for {\it dark matter} sector field imprints on the astrophysical phenomena are one of the most active branches of the current researches. Using numerical methods we elaborate the influence of {\it dark matter} on the emergence of black hole hair and formation of boson stars. We explore thermodynamics of different states of the system in Einstein-Maxwell-scalar {\it dark matter} theory with box boundary conditions.
Finally we find that the presence of dark sector within the system diminishes a chance of formation of scalar hair around a black hole.
\end{abstract}


\maketitle

\section{Introduction}
The astrophysical evidence of the illusive ingredient of our Universe, {\it dark matter}, is overwhelming and authorizes
the galaxy rotation curves, gravitational lensing, a thread-like structure (cosmic web) on which ordinary matter accumulates \cite{mas07, die12}.
On the contrary, the absence of the evidence of the most popular particle candidates for baryonic {\it dark matter} stipulates the necessity of diversifying 
experimental efforts \cite{ber18}. Black holes and ultra-compact horizonless objects being the ideal laboratories for {\it dark matter} studies, may help us to 
answer the tantalizing question of how {\it dark matter } sector leaves its imprint in the physics of these objects. However,
it happens that Schwarzschild black hole has a negative specific heat and it cannot be in equilibrium with thermal radiation. To overcome this difficulty the idea
of enclosing the black hole within a box was proposed \cite{yor86, bra90}. Einstein-Maxwell systems with box boundary conditions were elaborated in \cite{gib76}, where it was
established that the phase structure of the models were similar to AdS gravity. Inclusion of the additional scalar field to the theory in question, envisages the correspondence
of phase transitions in gravity in a box with s-wave holographic superconductor \cite{har08}-\cite{pen17}. The thermodynamical studies of Einstein-Maxwell scalar systems
in the asymptotically flat spacetime with reflecting boundary conditions were conducted in \cite{bas16}. A certain range of parameters allows to obtain stable black 
hole solution, giving a way to circumvent no-hair theorem.

The next compact objects studied in our paper, from the point of view of the influence of {\it dark matter} on their physics, are boson stars.
Boson stars being a self-gravitating solution of massive scalar field with a potential coupled to gauge fields and gravity \cite{jet92} are 
widely studied in literature \cite{kle09}-\cite{pen19}, for a quite long period of time.


The purpose of our paper is to examine thermodynamical properties and stability of the black holes and horizonless objects-boson stars in
Einstein-Maxwell-scalar system influenced by {\it dark matter} sector and envisage the role of the {\it dark matter} in the elaborated problems.

The organization of the paper is as follows. In Sec. II we describe the basic features of the {\it hidden sector} model and derived the basic equations needed in what follows.
Sec. III is devoted to the description of the obtained numerical results. In Sec. IV we concluded our researches.

\section{Model}
We consider the spacetime manifold with time-like boundary $\p \cM$, which will be referred as a box.
The action for Einstein-Maxwell scalar {\it dark matter} gravity is provided by
\begin{eqnarray} \label{ac}
S &=& \int_\cM d^4x\sqrt{-g} \Big(R - \frac{1}{4}F_{\mu\nu}F^{\mu\nu} - \frac{\alpha}{4}B_{\mu\nu}F^{\mu\nu} \\
&-& \frac{1}{4}B_{\mu\nu}B^{\mu\nu} - \vert D\Psi \vert^2 - m^2\vert\Psi\vert^2 \Big) - \int_{\partial \cM} d^3x \sqrt{-\gamma} \mathcal{K}, \nonumber
\end{eqnarray}
where $F_{\mu\nu}$ is a Maxwell field strength tensor, $B_{\mu\nu}$ is a strength tensor of a {\it hidden sector} vector boson. The complex scalar field 
$\Psi = \psi e^{i \theta}$, where $\theta$ denotes the phase,
is coupled only to the ordinary electromagnetic field by the covariant derivative $D_\mu = \nabla_\mu - iqA_\mu$.
The theoretical justifications of the model in question originate from M/string theories, where such mixing portals coupling Maxwell and auxiliary gauge fields 
can be encountered \cite{ach16}. The {\it hidden sectors} states are charged under their own groups and interact with the {\it visible} sector via gravitational interactions.
The realistic string compactifications establish the range of values for $\alpha$ between $10^{-2}$ to $10^{-10}$ \cite{abe04}-\cite{ban17}.
It seems that astrophysical observations of gamma rays of energy $511 keV$ \cite{jea03}, positron excess in galaxies \cite{cha08}, and muon anomalous magnetic moment
\cite{adr09}, argue for
the aforementioned idea of coupling Maxwell field with {\it dark matter} sector.
Recent experiments aimed at
gamma rays emissions from dwarf galaxies \cite{ger15}, dilaton-like coupling to photons caused by ultra-light {\it dark matter} \cite{bod15}, oscillations of the fine structure constant \cite{til15},
revisions of the constraints on {\it dark photon} 1987A supernova emission \cite{cha17}, 
measurements of excitation of electrons in CCD-like detector \cite{sensei}, as well as, the examinations in $e^+e^-$ Earth colliders \cite{lee14}, give us some hints for the 
correctness of the proposed model. They and the future planned ballon d'essai will ameliorate the mass constraints on the {\it hidden sector}
particles, especially for {\it dark photons}.

The second integral denotes the Gibbons-Hawking boundary term of our box with $\gamma$ metric on the three-dimensional hypersurface $(r = r_b)$, with
the extrinsic curvature $\mathcal{K}$.

Varying the action (\ref{ac}) we get the equations of motion of the forms
\ben \label{a1}
\Big( \na_\mu - iqA_\mu\Big) \Big(  \na^\mu &-& iqA^\mu\Big) \Psi - m^2 \Psi = 0,\\ \label{a2}
\talpha~\na_\mu F^{\mu \nu} &=& j^\nu,
\een
where $\talpha = 1 - \frac{\alpha^2}{4}$ and the current $j^\nu$ is provided by the relation
\be
j^\nu = iq \Big[ \Psi^\dagger \Big( \na^\nu - iqA^\nu\Big) \Psi - \Psi \Big( \na^\nu + iqA^\nu\Big) \Psi^\dagger \Big].
\ee

In what follows we use a time independent spherically symmetric line element, with the metric coefficients being functions of $r$-coordinate
\begin{equation}
ds^2 = - g(r)h(r)dt^2 + \frac{dr^2}{g(r)} + r^2(d\theta^2 + \sin\theta^2 d\phi^2),
\end{equation}
and the adequate components of the fields in the theory will constitute radial functions of the forms
\ben
A_\mu dx^\mu = \phi(r) dt,~B_\mu dx^\mu = \chi(r) dt, ~
\Psi = \Psi(r). 
\een
In general the scalar field can have harmonic time dependence which can be absorbed by a redefinition of the gauge field function. Having this in mind
it can be seen that the $r$-component of the equations of motion for the gauge and scalar fields leads the conclusion that $\Psi(r) = \psi(r)$. By virtue of this,  
the following equations of motion are provided:
\begin{eqnarray}
R_{\mu \nu} &-& \frac{1}{2} g_{\mu \nu} R = T_{\mu \nu}, \\
\nabla_\mu \nabla^\mu \psi &-& q^2 A_\mu A^\mu \psi - m^2 \psi = 0, \\
\nabla_\mu F^{\mu \nu} &+& \frac{\alpha}{2} \nabla_\mu B^{\mu \nu} - 2 q^2 A^\nu \psi^2 = 0, \\
\nabla_\mu B^{\mu \nu} &+& \frac{\alpha}{2} \nabla_\mu F^{\mu \nu} = 0.
\end{eqnarray}

As in the case of the equation (\ref{a2}), the last two equations can be rewritten as
\begin{eqnarray}
\talpha \nabla_\mu F^{\mu \nu} - 2 q^2 A^\nu \psi^2 = 0, \\
\nabla_\mu B^{\mu \nu} + \frac{\alpha}{\talpha} q^2 A^\nu \psi^2 = 0.
\end{eqnarray}

Consequently, the explicit forms of the equations of motion yield
\begin{eqnarray} \label{eq1}
h' &-& rh\psi'^2 - \frac{q^2 r \phi^2 \psi^2}{g^2} = 0, \\
g' &+& g\left(\frac{1}{r} + \frac{1}{2}r\psi'^2\right) + \frac{q^2 r \phi^2 \psi^2}{2gh} - \frac{1}{r} \nonumber \\
&+& \frac{r}{2h}(\phi'^2 + \alpha\chi' \phi' + \chi'^2 + m^2 h \psi^2) = 0,  \\
\phi'' &+& \left(\frac{2}{r} - \frac{h'}{2h}\right)\phi' - \frac{2 q^2 \phi \psi^2}{\talpha g} = 0, \\
\psi'' &+& \left(\frac{2}{r} + \frac{h'}{2h} + \frac{g'}{g}\right)\psi' + \left(\frac{q^2 \phi^2}{g h} - m^2\right)\frac{\psi}{g} = 0, \\ \label{eq5}
\chi'' &+& \left(\frac{2}{r} - \frac{h'}{2h}\right)\chi' + \frac{\alpha q^2 \chi \psi^2}{\talpha g} = 0.
\end{eqnarray}
To solve the equations of the theory in question one has to provide adequate boundary conditions. Namely we can pick either a horizonless or a black hole solution. 
In case of a black hole we expand the underlying functions in a Taylor series around the horizon of radius $r_h$
\begin{eqnarray} \label{psi1}
\psi &=& \psi_{0} + \psi_{1} (r-r_h) + \psi_{2} (r-r_h)^2 + \mathcal{O}(r^3), \\
\phi &=& \phi_{1} (r-r_h) + \phi_{2} (r-r_h)^2 + \mathcal{O}(r^3), \\
g &=& g_{1} (r - r_h) + g_{2} (r - r_h)^2 + \mathcal{O}(r^3), \\
h &=& 1 + h_{1} (r-r_h) + \mathcal{O}(r^2), \\ \label{chi1}
\chi &=& \chi_{1} (r-r_h) + \chi_{2} (r-r_h)^2 + \mathcal{O}(r^3).
\end{eqnarray}
We set $g_{0} = 0$, due to occurrence of the black hole
event horizon. For the regularity of the $U(1)$-gauge fields on the event horizon, one also puts 
 $\phi_{0}$ and $\chi_{0}$ equal to zero (in order to keep the terms with division by $g(r_h)$ in equations of motion finite).
By implementing the expansions (\ref{psi1})-(\ref{chi1})
into the equations of motion, we find out that $\{r_h, \psi_{0}, \phi_{1}, \chi_{1}, \alpha\}$ comprise free parameters of the theory in question, while the remaining ones can be expressed by them.

As far as the boson star scenario is concerned, we perform a similar expansion. However since the configuration in question is horizonless, 
the expansion accomplishes around the origin of the reference frame. At $r=0$ we require that the derivatives of all the functions are set equal to zero,
which ensures that there is no kink at  this point. At
$r=r_b$, we establish the Dirichlet boundary condition for the scalar field $\psi(r_b)=0$ (the reflecting mirror-like boundary conditions).


Asymptotic analysis of matter fields, at the box boundary, enables us to write 
\begin{eqnarray}
\psi \sim \psi^{(0)} + \psi^{(1)}(r_b - r) + \mathcal{O}(r^2), \\
\phi \sim \phi^{(0)} + \phi^{(1)}(r_b - r) + \mathcal{O}(r^2),\\
\chi \sim \chi^{(0)} + \chi^{(1)}(r_b - r) + \mathcal{O}(r^2).
\end{eqnarray}

As was proposed in Refs. \cite{bas16, pen19},
because of the fact that the scalar field satisfies the reflecting mirror-like boundary conditions $\psi(r_b)=0$, one can 
fix $\psi^{(0)} = 0$ and the other parameter $\psi^{(1)}$ can be used for the phase transition description. This approach to the problem in question
is widely exploited in holographic studies of superconductors and superfluids.

For the gauge fields one has that $\phi^{(0)} = \mu$ and $\chi^{(0)} = \mu_d$ as chemical potentials for {\it visible} and {\it hidden} sector fields, treating the system as a grand canonical ensemble.
In order to conduct the thermodynamical analysis we calculate the free energy of the system, to see which phase is thermodynamically preferable, for a fixed temperature. 
In the case of a hairless solution we take into account the classical formula $F = E - TS - \mu Q- \mu_d Q_d$, where $E$ is Brown-York quasilocal energy \cite{yor86,bra90}. 
Nevertheless this approach may cause problems in hairy solution analysis, being insufficient to capture the mass of the scalar field. Therefore 
we treat this problem evaluating the on-shell action in Euclidean signature $F = TS_{cl}$, which enables to take into considerations the non-trivial
profile of scalar field constituting the solution of the underlying system of differential equations.

Solution of the equations (\ref{eq1})-(\ref{eq5}) with $\psi = 0$ can be achieved analytically, giving the Reissner-Nordstrom (RN) {\it dark matter} black object \cite{kic19}.
To proceed further and accomplish the complete numerical analysis of the underlying equations, we implement the shooting method, integrating the aforementioned
relations from $r_h$ to $r_b$,
using the fourth order Runge-Kutta method. 
From the set of free parameters we fix the scalar magnitude on the event horizon $\psi_{0}$ and pick $r_h$, $\phi_{1}$ and $\chi_{1}$ to be shooting parameters.
Moreover we impose values on both chemical potentials, that serve as constrains in our shooting procedure for $\phi_{1}$ and $\chi_{1}$.
We set a domain of shooting parameters from the series expansion of the solutions near the horizon, then by using the iterative bisections
 one finds a solution that meets constrains, with a desired tolerance.
Therefore parameters $\{ \mu_{d}, ~\alpha \}$, which are controlling respectively amount of \emph{dark charge} and the coupling strength remain free, thus they can be varied to see their impact on the system in question.
For the convenience let us refer to the parameter $\psi^{(1)}$, as a \textit{condensation}, which serves as a handy analogy to holographic theory.
As mentioned above, in our numerical scheme we treat $\mu_{d}$ as an input parameter in our code, however one might not be interested in expressing these relations in a language of chemical potentials. Therefore one might compute the total \emph{dark charge} of the system
\begin{equation}
Q_d = \lim_{r \rightarrow r_b} \frac{1}{4\pi} \int_{\mathcal{S}^2} B_{\mu \nu} t^{\mu} n^{\nu} \sqrt{-g} d^2\theta,
\end{equation}
where $t_{\mu}$ is a unit time-like vector and $n_{\mu}$ is a normal vector to the boundary.
In the similar manner we compute electrical charge, for $F_{\mu \nu}$.

\section{Results}

\begin{figure}[h]
\centering
\includegraphics[width=0.5\textwidth]{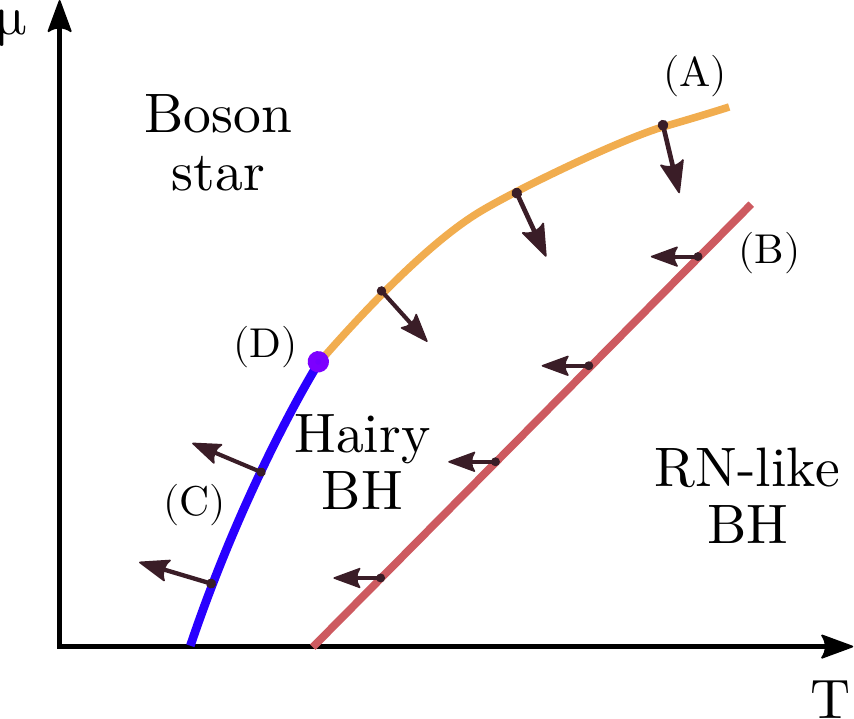}
\caption{A scheme of phase diagram of the described system. Blue-yellow line indicates the border between boson star and hairy black hole parameter space, while the red line depicts hairy BH -- generalized RN BH phase boundary.
The arrows on the scheme show us the flow of phase boundaries driven by the chemical potential of \textit{dark matter}. Lines have been split and labeled from A to C, with a point D being the center of rotation of left-hand side boundary.}
\label{fig:phase_diagram}
\end{figure}

We commence with the hairy black hole solution (HBH), i.e., a system with an event horizon and non-trivial scalar field profile.
The parameter space of HBH can be illustrated on a plane of chemical potential and Hawking temperature ($\mu$--$T$) as a triangular shape.
That region is bound between boson star phase from the left-hand side and generalized Reissner-Nordstrom (RN) solution from the right-hand side.
A schematic phase diagram has been presented in the Fig. \ref{fig:phase_diagram}, where both mentioned lines are marked.
Moreover the influence of the \textit{dark sector} on phase boundaries is visualized by arrows, showing the trend of the flow by increasing the hidden sector chemical potential.

\begin{figure}
\centering
\includegraphics[width=0.5\textwidth]{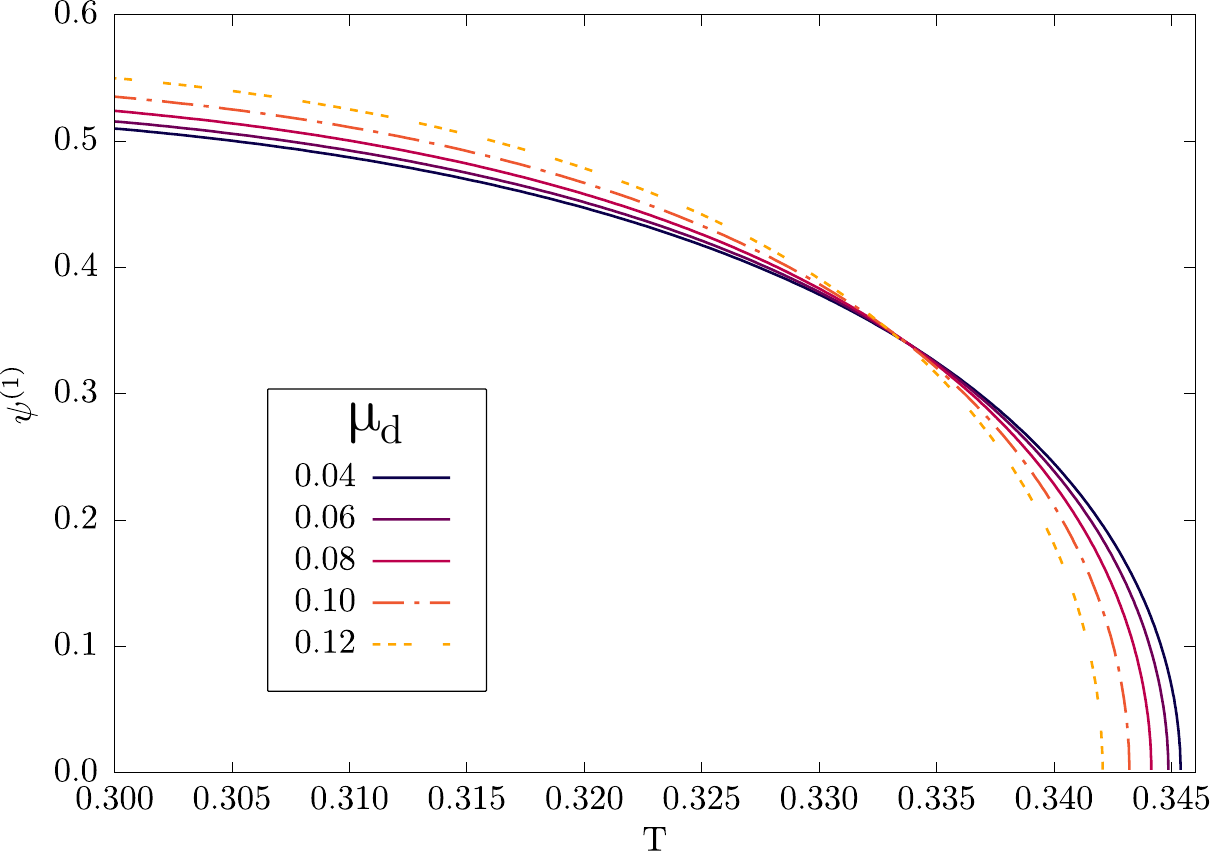}
\caption{Condensation $\psi^{(1)}$ as a function of temperature, for the different values of $\mu_{d}$ and $\alpha = 10^{-3}$.
For $\mu=0.1$ a typical second order phase transition takes place, the \textit{dark matter} presence influences the transition point and the condensation.}
\label{fig:hbh_normtrans}
\end{figure}

The hairy configuration can be achieved for a specific value of the chemical potential. Below the value $\mu_{RN}$ scalar can not condensate and we 
get RN-{\it dark matter} black hole. On the other hand, for the value greater than the critical one, $\mu_c$, the system becomes unstable.
By stable hairy solution we mean a constrained solution of the equations of motion \eqref{eq1}-\eqref{eq5} that fulfils the boundary conditions with desired tolerance and its free energy is lower than the free energy of RN and BS, making it the ground state of the system.
We can define $\mu_c$ as the chemical potential for which the phase transition driven by temperature is no longer of second order and the condensate collapses.
In the range between $\mu_{RN}$ and $\mu_c$, we contend a typical second order phase transition, depicted in Fig. \ref{fig:hbh_normtrans}.
In the vicinity of critical temperature the {\it condensation} can be described by a function \BK{$\psi^{(1)} \sim (T_c - T)^{1/2}$}.
It should also be noted that establishing a HBH solution requires relatively large value of the scalar charge. In our calculation we used $q=100$
and a small mass of $m=10^{-6}$.

\begin{figure}
\centering
\includegraphics[width=0.5\textwidth]{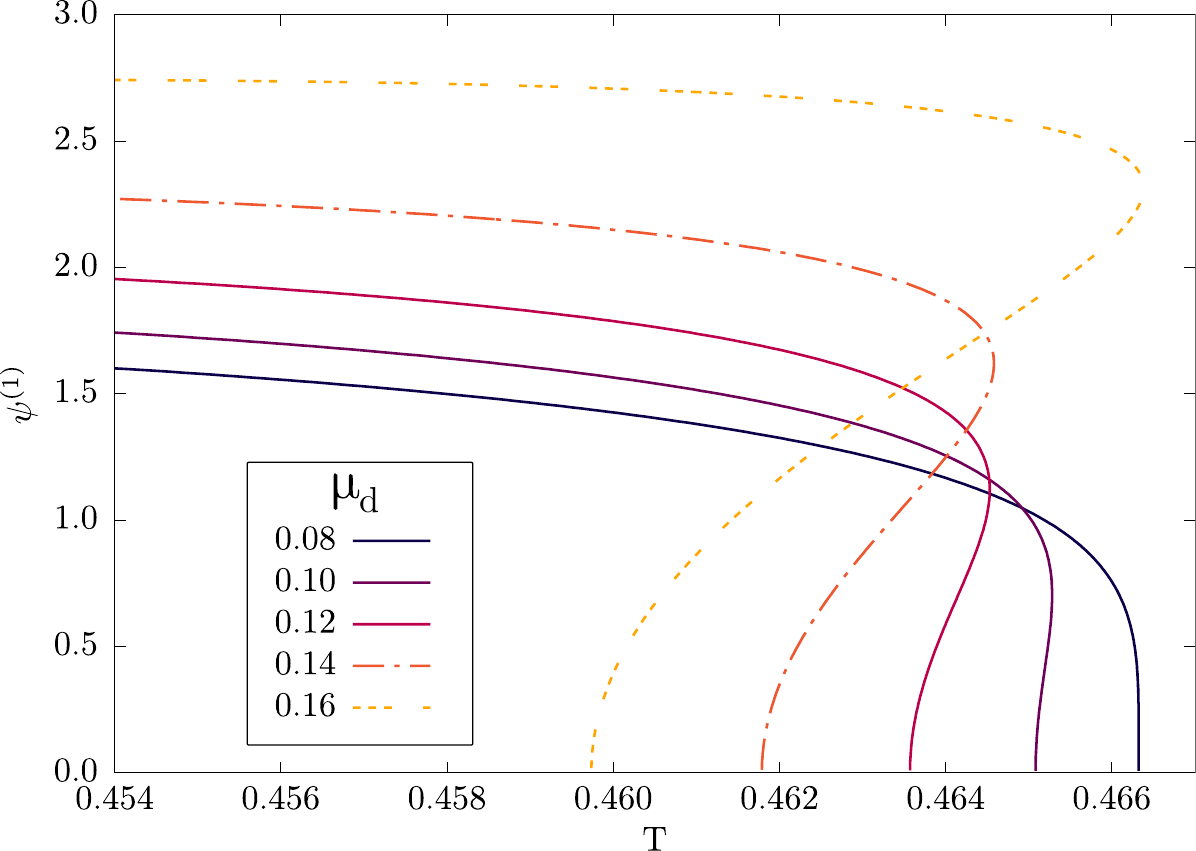}
\caption{Double valued profiles of the \textit{condensation} as functions of Hawking temperature caused by increasing amount of \textit{dark matter} in the system with $\alpha = 10^{-3}$ and $\mu = 0.14$. While the first transition for $\mu_{d} = 0.08$ might still be considered as a regular the another strictly not -- the value of condensation becomes double valued for some range of temperatures.
Moreover these solutions obey the boundary conditions but their free energy is larger than both BS's and RN's, therefore they cannot be considered as thermodynamically preferred.}
\label{fig:hbh_exoticphase}
\end{figure}

Let us now discuss physical mechanisms behind the phase boundaries flow from the Fig. \ref{fig:phase_diagram}.
When we cross the line of the critical chemical potential value $\mu_c$, one encounters the {\it exotic phase}, where for one value of temperature we have two values of the {\it condensation} parameter $\psi^{(1)}$.
Moreover by evaluating its free energy we can find it is so high that the hairy state is no longer stable -- our numerical method finds constrained solutions, but due to free
energy leap, they are not thermodynamically preferred.
The \emph{exotic phase} effect occurs in case when scalar mass is close to or equal zero, 
for a mass away from this limit we do not obtain that phase.
Instead we have a sharp crossing, from stable solutions below $\mu_c$ to the situation when the equations of motion do not provide solutions with condensed scalar above the 
$\mu_c$ threshold at all.
It is worth mentioning that a similar condensation--temperature profile has been shown in the so-called vector p-wave holographic superfluids \cite{cai14}, but a first order phase transition was hidden behind it. However,
it was revealed that for the real value of the vector field, the model in question gave us the same description as holographic s-wave model with {\it dark matter}
sector \cite{rog16}-\cite{rog16a}.

{\it Dark matter} gauge field plays an interesting role in this transition, as it accelerates the appearance, let us say,  the \textit{exotic} phase.
For a larger value of \textit{dark sector} chemical potential, $\psi^{(1)}$ becomes double valued for the lower chemical potential, which is depicted in the Fig. \ref{fig:hbh_exoticphase}, where $\mu_d$ has gradually increasing value.
Moreover when system enters the \emph{exotic} phase its free energy rises repeatedly and exceeds the free energy of RN black hole, so the hairy phase 
is no longer a preferred option.
In this way, that effect restricts the range of chemical potential where the second order phase transition may occur, $\mu_c$ becomes a descending 
function of the \textit{dark charge} (see curve (A) in the Fig. \ref{fig:phase_diagram}).
However it cannot be increased without a limit.
For every value of the electric charge there exists a certain limit of \textit{dark charge}, below which a formation of scalar hair is possible.
Above it, such condensation cannot take place and no stable solutions are found.
This phenomenon adds up both gravitational influence of the charge on the metric and non-gravitational coupling between both gauge fields.

\begin{figure}
\centering
\includegraphics[width=0.5\textwidth]{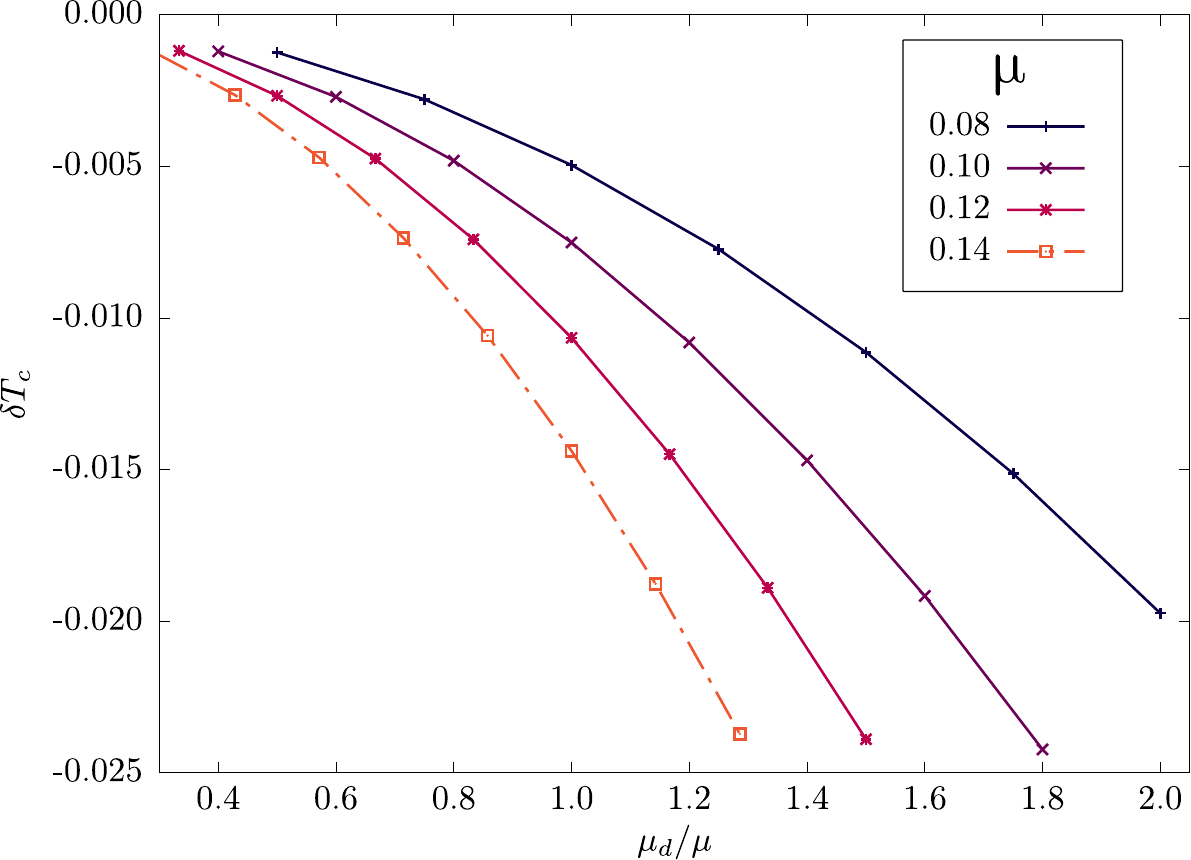}
\caption{Relative change of the critical temperature of hairy black hole -- generalized RN black hole as a function of a ratio of chemical potentials.
The temperature ratio has been normalized to the critical temperature of \textit{dark matter} free solution, where $\mu_{d} = 0$.}
\label{fig:tc_shift}
\end{figure}

Now let us draw our attention to the HBH-RN BH (B) border.
An interesting effect that {\it hidden sector}
exerts on the hairy black hole system is the shifting of the critical temperature of the phase transition. The larger growth of {\it dark matter} charge (and also $\mu_{d}$) we observe, the lower value of the transition temperature one achieves.
Such effect has been depicted in the Fig. \ref{fig:tc_shift}, where the critical temperature ratio described by the relation
$$ \delta T_c = \frac{T_c(\mu_{d}) - T_c(0)}{T_c(0)}, $$
is shown as a function of the chemical potential of \textit{hidden sector} normalized to the \textit{visible sector} chemical potential. 
One can notice that the shift of the critical temperature is proportional to the square of $\mu_{d}$.
Obviously it can not decrease as low as one wishes and a certain limit exist which has been discussed in analytic solution of {\it dark matter }charged RN-like black hole \cite{kic19}.
The descent of the critical temperature becomes steeper for larger value of the chemical potential of the \textit{visible sector}.
It can be explained by the non-gravitational interaction between fields via \textit{kinetic mixing} term, which plays the significant role when both fields are sufficiently strong.


\begin{figure}
\centering
\includegraphics[width=0.5\textwidth]{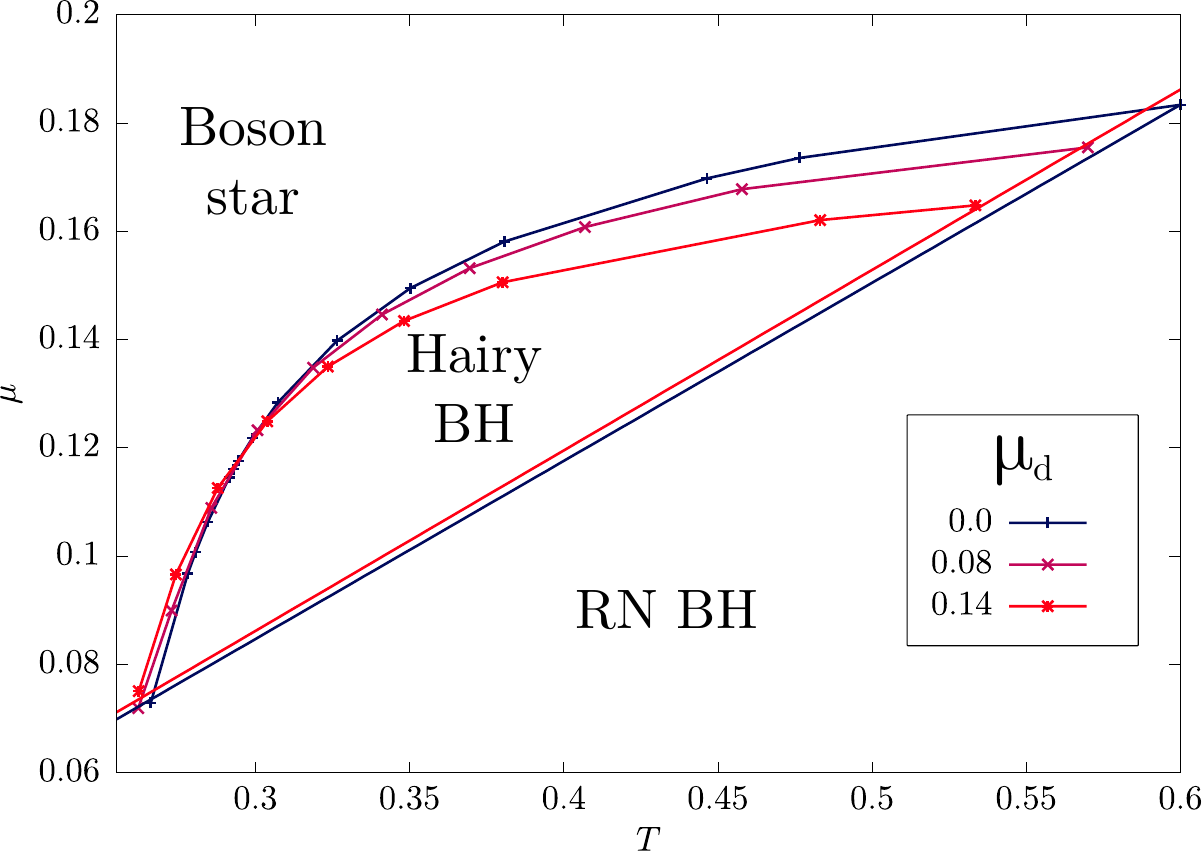}
\caption{Quasi-rotation of the phase transition boundary between boson star and hairy black hole caused by \textit{dark sector} charge with $\alpha$-coupling equal to $10^{-4}$.
It can be observed that the $\mu$ threshold for hairy BH solution is significantly lowered and some parameter space of boson star is taken for the advantage of hairy BH for lower values of the chemical potential.}
\label{fig:hbh_bs_phasediag}
\end{figure}

To proceed further, we shed some light on the influence of {\it dark matter} sector on the black hole-boson star phase transition in the stable area of small values of the chemical potential (line (C) in Fig. \ref{fig:phase_diagram}). This process is depicted in Fig. \ref{fig:hbh_bs_phasediag}, which asserts a phase diagram at the boundary between hairy black hole and boson star.
While the boson star is a horizonless object its Hawking temperature remains undefined. However it is possible to calculate its characteristic -- condensation and free energy as a function of chemical potential.  
Then to obtain the phase boundary curve, we start in the hairy black hole regime, then one moves towards lower Hawking temperatures and study the value of the free energy of the hairy black hole on the way. When it exceeds the free energy of a boson star, for the corresponding value of the chemical potential, the transition point is found. 
Both phases of the system are influenced by the \textit{hidden sector}, nonetheless the free energy of a boson star is affected much less than black hole's. \textit{Dark sector} causes a significant drop of free energy of a hairy black hole.
It means that the stability of a hairy solution is preserved for lower temperatures given the presence of the \textit{dark matter} in the system.
Such effect causes that hairy black hole solution is thermodynamically preferable for the lower Hawking temperatures and limits the emergence of boson star. The presence
of $\alpha$-coupling constant slightly diminishes the space of parameters for which boson star can emerge.


At last it is sensible to mention some points that seem to be \textit{dark sector} resistant.
One of them appears on the phase boundary, labeled with (D) on the phase diagram scheme in Fig. \ref{fig:phase_diagram}. This point or rather its neighborhood does not seem to be susceptible on the \textit{dark charge} presence in the system.
For a particular numerical example, like in the Fig. \ref{fig:hbh_bs_phasediag}, it is placed around $\mu \approx 0.1160942$ and $T \approx 0.2929537$.
Another one can be noticed in the condensation--Hawking temperature dependence presented in Fig. \ref{fig:hbh_normtrans}.
All the curves certainly cross each other in one point, located at $\psi^{(1)} \approx 0.345$ and $T \approx 0.3325$. This interesting phenomenon shows that while the \textit{dark sector} may modify the phase structure of the system and has an imprint on its critical quantities, there exists a specific configuration of the system, that remains completely untouched.

\begin{figure}
\includegraphics[width=0.5\textwidth]{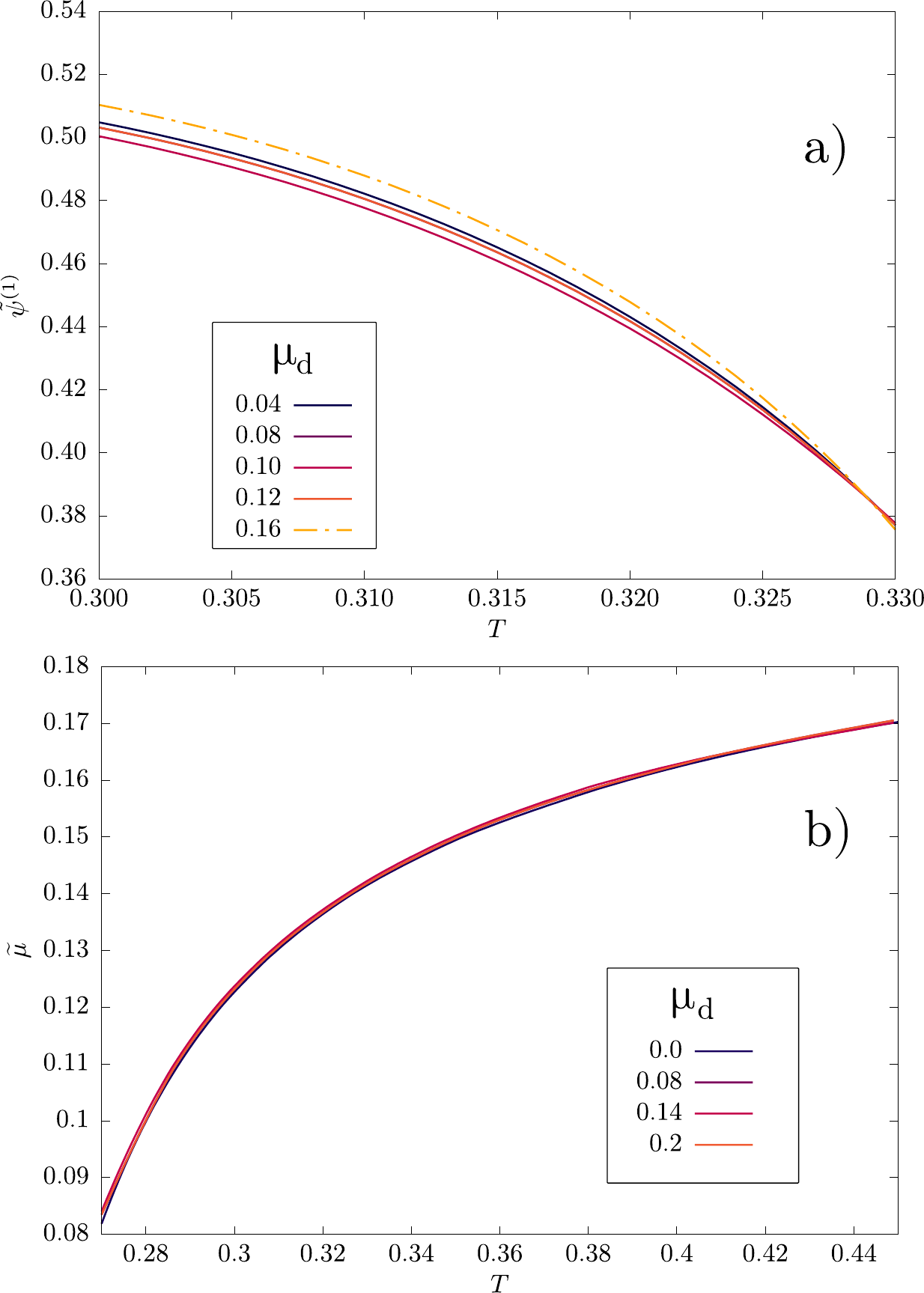}
\caption{
Panel a) presents condensation profile after the transformation performed in equation (\ref{eq:trans_dm}). The plot refers to the same data as in Fig. \ref{fig:hbh_normtrans}, however it can be seen that the separation between curves is significantly smaller.
In case of panel b), which refers to Fig. \ref{fig:hbh_bs_phasediag}, all curves appear to overlap with the \emph{dark matter} free solution. 
The aforementioned transformation had removed the leading term of the \emph{dark sector} influence.}
\label{fig:isosbest_transformation}
\end{figure}

The points in question constitute the so-called {\it isosbestic} ones \cite{isos}, where the curves dependent on temperature $T$ and parametrized by values of {\it dark matter}
chemical potential, intersect. They illustrate the influence of temperature on condensation $\psi^{(1)}$ and chemical potential of {\it visible sector}.
At this point we may perform a short analysis, which would reveal the leading order term of the \emph{dark sector} influence.
We take a following expansion of the condensate
\begin{equation}
\psi^{(1)}(T, \mu_{d}) = \psi^{(1)}(T, 0) + \mu_{d}^2 ~\psi^{(1)}_1(T) + \mathcal{O}(\mu_{d}^3).
\end{equation}
Second order term takes the approximated form
\begin{equation}
\psi^{(1)}_1 (T) = \frac{\psi^{(1)}(T, \mu_{d1}) - \psi^{(1)}(T, \mu_{d2})}{\mu_{d1}^2 - \mu_{d2}^2},
\label{ppp}
\end{equation}
where in a certain example of curves from Fig. \ref{fig:hbh_normtrans} we took $\mu_{d1} = 0.12$ and $\mu_{d2} = 0.08$.
The zero of this function refers to the isosbestic point, where the contribution of $\mu_d$ is by definition none.
By calculating above function with help the leading order of the influence of the \emph{dark sector} may be subtracted from the main function
\begin{equation}
{\tpsi}^{(1)}(T, \mu_{d}) = \psi^{(1)}(T, \mu_{d}) - \mu_{d}^2 ~\psi^{(1)}_1(T).
\label{eq:trans_dm}
\end{equation}
In the similar manner we can expand and analyze the chemical potential as a function of Hawking temperature, parametrized by $\mu_d$ from boson star -- hairy black hole phase boundary
\begin{equation}
\mu(T, \mu_d) = \mu(T, 0) + \mu_d^2 ~\mu_1(T) + \mathcal{O}(\mu_d^3).
\end{equation}
We define $\mu_1(T)$ analogically to \eqref{ppp} with $\mu_{d1} = 0.2$ and $\mu_{d2} = 0.08$ and perform the same transformation for $\mu(T, \mu_d)$ curve as for $\psi(T, \mu_d)$ in \eqref{eq:trans_dm}.
The effect of these transformations is depicted in Fig. \ref{fig:isosbest_transformation}, where all curves tend to be much closer to each other than before.
Obviously the total effect of $\mu_d$ is not ruled out completely, since it is much more complex than in the considered expansion.

\section{Conclusion}
In our paper, based on Einstein-Maxwell scalar {\it dark matter} theory, where the {\it hidden sector} is mimicked by 
the auxiliary $U(1)$-gauge field coupled to the ordinary Maxwell one by the {\it kinetic mixing term} with a coupling constant $\alpha$, we elaborate two
scenarios of emergence of a hairy black hole or a boson star.
The main motivation standing behind our research was to shed some light on the influence of {\it dark matter} sector on the physics and thermodynamics of these systems.

The obtained results reveal that the coupling between {\it visible} and {\it hidden} sectors plays a complex role in the behavior of scalar hair. 
The parameter space ($\mu - T$), where these solutions constitute a thermodynamically favorable phase, is being narrowed on two boundaries and 
extended to another one.
The {\it dark sector}'s presence strongly reduces the value of critical chemical potential, above which the hairy solution becomes unstable. Moreover the critical temperature of HBH--RN-like solution is shifted towards lower value of Hawking temperature.
However the boundary between HBH and boson stars is shifted towards the latter.
The presence of the {\it dark sector} lowers the free energy of HBH system, which 
broadens the parameter space available for the emergence of the object in question by a noticeable extent, i.e., leaving boson star as an adverse phase in the low $\mu$ regime.
It appears that the free energy of boson stars in the considered configuration reacts faintly to the presence of $U(1)$-gauge {\it dark matter} field.
While the response of the system is visible it is much smaller in magnitude than of the condensate around a black hole.
However we suppose that interesting results may be achieved for more robust model of a scalar field, e.g., containing self interacting terms.

In the view of presented results it seems that the hairy solutions are not only battled by no-hair theorems originating from the theory of black holes, but also by a factor that is commonly present in our universe -- the {\it dark matter}.
Even if such formation would be possible despite different obstacles a significant abundance of {\it dark matter} may prevent hairy solutions from emerging.

To visualize the impact of dark sector, we compute the area of HBH parameter space between both phase boundaries.
One can consider simple integration $\int (\mu_{BS}(T) - \mu_{RN}(T))dT$ of the curves from  the Fig. \ref{fig:hbh_bs_phasediag}, which 
reveals that the \textit{dark sector} with $\mu_{d} = 0.14$ takes away approximately $27\%$ of the hairy black hole's parameter space, compared to {\it dark matter} free solution.
It is indeed a significant difference, because even if such formation would be possible despite different obstacles a significant abundance of \textit{dark sector} may prevent hairy solutions from emerging.

The curves $\psi^{(1)}(T)$ and $\mu(T)$, parametrized by the values of {\it dark matter} chemical potential reveal the {\it isosbestic points}, where they all intersect.
One has the specific configurations of the considered system which is unaffected by the influence of {\it hidden sector}.
At the points in question we perform analysis revealing that the leading order influence of {\it dark matter} on the condensation $\psi^{(1)}$ and chemical potential
of ordinary matter is quadratic in $\mu_d$.

As a concluding remark, we present promising future research directions. We have elaborated the simple box-boundary models of a hairy 
black hole and a boson star (the so-called small boson star).
The tantalizing question can be asked about the different boson star configurations with additional fields and potentials. Further investigations in this direction will be 
published elsewhere.

\begin{acknowledgments}
We acknowledge K. I. Wysokinski and N. Sedlmayr for fruitful discussions on various occasions.
\end{acknowledgments}


\end{document}